\documentclass{revtex4}

\usepackage{amsmath}
\usepackage{graphicx}
\usepackage{subfigure}
\usepackage{xspace}
\usepackage{algorithm}
\usepackage{algorithmic}
\usepackage{listings}
\newcommand{\CXX}{{C\nolinebreak[4]\hspace{-.05em}\raisebox{.3ex}{\tiny\bf ++}}\xspace}
\newcommand{\code}[1]{\lstinline|#1|}

\begin{document}

\title{AMGCL: an Efficient, Flexible, and Extensible\\ Algebraic Multigrid Implementation}

\author{\firstname{Denis}~\surname{Demidov}}
\email[E-mail: ]{dennis.demidov@gmail.com}
\affiliation{
    Kazan Branch of Joint Supercomputer Center, Scientific Research Institute of System Analysis,
    the Russian Academy of Sciences, Lobachevsky st. 2/31, 420111 Kazan, Russian Federation}

\begin{abstract}
    The paper presents AMGCL~-- an opensource \CXX library implementing the
    algebraic multigrid method (AMG) for solution of large sparse linear
    systems of equations, usually arising from discretization of partial
    differential equations on an unstructured grid. The library supports both
    shared and distributed memory computation, allows to utilize modern
    massively parallel processors via OpenMP, OpenCL, or CUDA technologies, has
    minimal dependencies, and is easily extensible. The design principles
    behind AMGCL are discussed and it is shown that the code performance is on
    par with alternative implementations.
\end{abstract}


\maketitle

\section{Introduction}

The ability to solve large sparse linear system of equations that arise from
discretizations of partial differential equations on either structured or
unstructured grids is extremely important in modern numerical methods. Direct
methods fail to scale beyond a certain size, typically of the order of a few
millions of unknowns~\cite{hogg2013new, henon2002pastix}, due to their
intrinsic memory requirements and shear computational cost.  This makes
preconditioned iterative methods the only viable approach for solution of
large scale problems.

The combination of a Krylov subspace method \cite{saad2003iterative} with
algebraic multigrid (AMG) as a preconditioner is considered to be one of the
most effective choices for solution of such systems~\cite{brandt1985algebraic,
ruge1987algebraic, Trottenberg2001}.  The AMG can be used as a black box solver
for various computational problems, since it does not require any information
about the underlying geometry, and is known to be robust and scalable
\cite{cleary2000robustness}.  There are several well-known AMG implementations
available today. Notable examples are Trilinos~ML package~\cite{ml-guide},
BoomerAMG from Hypre~\cite{falgout2002hypre}, and GAMG from
PETSC~\cite{petsc-user-ref}. These packages are provided as parts of complex
frameworks targeting large distributed memory machines, have a steep learning
curve in general, and may be difficult to compile or distribute.  Another
problem is that most of the available packages are licensed under GPL, and so
are usually deemed inappropriate for use in commercial software. Being as large
and inert as they are, the frameworks are usually slow to implement support for
modern hardware, such as CUDA of OpenCL based GPUs. There are smaller packages
that address this issue. For example, the CUSP library provides implementation
of smoothed aggregation multigrid~\cite{Cusp, bell2012cuspamg}, but only
supports NVIDIA hardware. The ViennaCL compute library~\cite{ViennaCL} is
another example of GPGPU (general purpose GPU) AMG implementation, which also
supports OpenMP and OpenCL standards. However, neither of the GPGPU libraries
provide support for distributed memory clusters.

The AMGCL \CXX library (published at https://github.com/ddemidov/amgcl)
provides an AMG implementation while trying to address the above issues. It has
minimal set of dependencies, is published under permissive MIT license, targets
both shared and distributed memory machines, and supports modern many-core
architectures.  The library allows to utilize user-defined data structures and
operations, thus making it easy to integrate AMGCL into existing software with
large and stable codebase. This paper serves as an introduction to the library,
discusses its design principles, and compares performance of the provided
algorithms with alternative implementations.

\section{AMG overview}

This section outlines the basic principles behind the algebraic multigrid
method~\cite{brandt1985algebraic, Stuben1999}. Consider a system of linear
algebraic equations in the form
\begin{equation} \label{eq:auf}
    Au = f,
\end{equation}
where $A$ is a square matrix. Mutigrid methods are based on recursive
application of a two-grid scheme, which combines \emph{relaxation} and
\emph{coarse grid correction}. Relaxation, or smoothing iteration, is a simple
iterative method, such as Jacobi or Gauss--Seidel
iteration~\cite{barrett1994templates}. Coarse grid correction solves the
residual equation on a coarser grid, and improves the fine-grid approximation
with the interpolated coarse-grid solution. Transfer between grids is described
with \emph{transfer operators} $P$ (\emph{prolongation} or
\emph{interpolation}) and $R$ (\emph{restriction}).

In geometric multigrid methods the matrices $A_i$ and operators $P_i$ and $R_i$
are usually supplied by the user based on the problem geometry. In algebraic
multigrid methods the grid hierarchy and the transfer operators are in general
constructed automatically, based only on the algebraic properties of the
matrix~$A$. The \emph{setup} phase of a generic AMG algorithm may be described
as follows:
\begin{algorithm}[H]
\caption{AMG setup}
\begin{algorithmic}
    \STATE Start with a system matrix $A_1 \leftarrow A$.
    \WHILE{the matrix $A_i$ is too big to be solved directly}
        \STATE Introduce prolongation operator $P_i$ and restriction
            operator $R_i$.
        \STATE Construct the coarse system using Galerkin operator: $A_{i+1}
            \leftarrow R_i A_i P_i$.
    \ENDWHILE
    \STATE Construct a direct solver for the coarsest system $A_L$.
\end{algorithmic}
\end{algorithm}

Note that in order to construct the next level in the AMG hierarchy, one only
needs to define transfer operators $P$ and $R$. Also, a common choice for the
restriction operator $R$ is a transpose of the prolongation operator: $R=P^T$.
After the AMG hierarchy is constructed, it is used to solve the
system~\eqref{eq:auf} as follows:
\begin{algorithm}[H]
\caption{AMG solve}
\begin{algorithmic}
    \STATE Start at the finest level with initial approximation $u_1$.
    \WHILE {not converged}
        \STATE \COMMENT{ \emph{V-cycle:} }
        \FOR {each level of the hierarchy, finest-to-coarsest}
            \STATE Apply a couple of smoothing iterations
                (\emph{pre-relaxation}) to the current solution:
            \STATE $u_i \leftarrow S_i(A_i, f_i, u_i)$
            \STATE Find residual $e_i \leftarrow f_i - A_i u_i$ and restrict it
                to the right-hand side on the coarser level:
            \STATE $f_{i+1} \leftarrow R_i e_i$
        \ENDFOR
        \STATE Solve the coarsest system directly: $u_L \leftarrow A_L^{-1} f_L$.
        \FOR {each level of the hierarchy, coarsest-to-finest}
            \STATE Update the current approximation with the
                interpolated solution from the coarser level:
            \STATE $u_i \leftarrow u_i + P_i u_{i+1}$
            \STATE Apply a couple of smoothing iterations
                (\emph{post-relaxation}) to the updated solution:
            \STATE $u_i \leftarrow S_i(A_i, f_i, u_i)$
        \ENDFOR
    \ENDWHILE
\end{algorithmic}
\end{algorithm}

Usually AMG is not used standalone, but as a preconditioner with a Krylov
subspace iterative solver. In this case single V-cycle is used as a
preconditioning step.

\section{AMGCL design principles}

The most algorithmically challenging part of the AMG is the setup phase, or,
more specifically, finding a set of transfer operators that would work well for
the problem at hand. Most of the approaches here are intrinsically serial and
are hard to parallelize.  On the contrary, once the hierarchy is constructed,
the solution step is mostly trivial and may be expressed through several
well-defined and easily parallelizable primitives such as sparse matrix-vector
product or linear vector combinations. In terms of machine time, the setup
usually takes minor part of the overall algorithm. Usually this is no more
than~30\%, often much smaller, depending on the convergence rate of the
problem.  Also, since the setup phase only depends on the system matrix, it is
possible to reuse the constructed hierarchy with multiple right-hand sides.
This is especially important for solution of nonstationary
problems~\cite{Demidov2012}.

This difference between algorithmic and computational complexity of the setup
and the solution phases of the AMG method is the reason for the most important
design decision behind the AMGCL library. Namely, the AMG hierarchy is always
constructed on the main processor (CPU), using the internal data structures and
employing the OpenMP parallelization standard where appropriate. The
constructed hierarchy is then transferred into one of the provided (or
user-defined) backends for the solution phase. The solution step, depending on
the selected backend, may utilize various parallelization technologies, such as
OpenMP~\cite{dagum1998openmp}, OpenCL~\cite{stone2010opencl}, or
CUDA~\cite{sanders2010cuda}.

AMGCL is implemented as a header-only \CXX library, and uses the compile-time
policy-based design~\cite{alexandrescu2001modern}, which allows its users to
compose their own customized version of the AMG from the algorithmic components
described below.

\begin{description}
    \item[Backends.] Backend is a class that defines matrix and vector types
        and defined parallel primitives that are utilized during the solution
        phase of the algorithm. This level of abstraction enables transparent
        acceleration of the solution with OpenMP, OpenCL, or CUDA technologies,
        and also allows users of the library to employ their own data
        structures either to avoid unnecessary data copies or to use faster
        algorithms (for example, the ones that are customized for a specific
        hardware).
    \item[Value types.] Value types allow to generalize AMGCL algorithms onto
        complex or non-scalar systems. A value type defines a number of
        overloads for common math operations, and is used as a template
        parameter for a backend. Most often, a value type is simply a builtin
        \code{double} or \code{float}, but it is also possible to use small
        statically sized matrices as value type, which may increase cache
        locality, or convergence ratio, or both, when the system matrix has a
        block structure.
    \item[Coarsening strategies.] These are various options for creating coarse
        levels in the AMG hierarchy. A coarsening strategy takes the system
        matrix $A$ at the current level, and returns prolongation operator $P$
        and the corresponding restriction operator $R$. AMGCL provides
        classical Ruge--Stuben strategy, smoothed and non-smoothed aggregation,
        and smoothed aggregation with energy
        minimization~\cite{ruge1987algebraic, vanvek1996algebraic,
        sala2008new}.
    \item[Relaxation methods.] These are the smoothers that are used on each
        level of the AMG hierarchy during solution phase and are expressed in
        terms of parallel primitives defined by the current backend. Supported
        options include damped Jacobi iteration, sparse approximate inverse,
        Gauss--Seidel, incomplete LU decomposition, and Chebyshev
        relaxation~\cite{barrett1994templates, saad2003iterative,
        broker2002sparse}.
    \item[Iterative solvers.] Krylov subspace methods that may be combined with
        the AMG preconditioner to solve the linear system~\eqref{eq:auf}.
        AMGCL provides implementation for several variants of CG,
        BiCGStab, GMRES, and IDR(s)~\cite{saad2003iterative,
        fokkema1996enhanced, baker2005technique, sonneveld2008idr}.
        Similarly to a smoother, an iterative solver is expressed in terms of
        the parallel primitives defined by the backend.
\end{description}

Listing~\ref{lst:example} shows an example of a solver declaration that uses a
BiCGStab iterative method preconditioned with a smoothed aggregation AMG. A
sparse approximate inverse method (SPAI-0) is used as a smoother on each level
of the hierarchy. The solver uses the bultin backend (accelerated with OpenMP)
and double precision scalars as value type.

\begin{lstlisting}[float=t,
    caption={An example of AMGCL solver declaration.}, label=lst:example]
#include <amgcl/backend/builtin.hpp>
#include <amgcl/make_solver.hpp>
#include <amgcl/amg.hpp>
#include <amgcl/coarsening/smoothed_aggregation.hpp>
#include <amgcl/relaxation/spai0.hpp>
#include <amgcl/solver/bicgstab.hpp>

typedef amgcl::backend::builtin<double> Backend;

typedef amgcl::make_solver<
    amgcl::amg<
        Backend,
        amgcl::coarsening::smoothed_aggregation,
        amgcl::relaxation::spai0
        >,
    amgcl::solver::bicgstab<Backend>
    > Solver;
\end{lstlisting}

\subsection{Algorithm parameters}

Each component in AMGCL defines its own parameters by declaring a \code{param}
subtype with some meaningful default values. When a class is composed from
several subclasses, it includes parameters of its children into its own
\code{param} struct.  This allows to provide a unified interface to the various
algorithmic components in AMGCL.  For example, parameters for the
\code{make_solver<Precond, Solver>} class from Listing~\ref{lst:example} are
declared as

\begin{lstlisting}
template <class Precond, class Solver>
struct make_solver {
    struct params {
        // Preconditioner parameters:
        typename Precond::params precond;
        // Iterative solver parameters:
        typename Solver::params solver;
    };
};
\end{lstlisting}

\subsection{Runtime interface}

The compile-time configuration of AMGCL solvers is efficient, but is not always
convenient, especially if the solvers are used inside a software package or
another library and it is not possible to choose an optimal configuration a
priori.  The runtime interface allows to postpone some of the configuration
decisions until the program is running and the type of problem being solved is
known. The classes inside \code{amgcl::runtime} namespace correspond to their
compile-time alternatives, but their only template parameter is the backend to
use.

Since there is no way to know the parameter structure at compile time, the
runtime classes accept parameters as instances of
\code{boost::property_tree::ptree} class provided as part of Boost.PropertyTree
library\footnote{https://www.boost.org}.  The actual components of the method
are set through the property tree as well.  For example,
Listing~\ref{lst:runtime} shows the runtime alternative to the solver declared
in Listing~\ref{lst:example}.

\begin{lstlisting}[float=t,
    caption={Runtime definition of an AMGCL solver.}, label=lst:runtime]
#include <amgcl/backend/builtin.hpp>
#include <amgcl/make_solver.hpp>
#include <amgcl/amg.hpp>
#include <amgcl/coarsening/runtime.hpp>
#include <amgcl/relaxation/runtime.hpp>
#include <amgcl/solver/runtime.hpp>

typedef amgcl::backend::builtin<double> Backend;

typedef amgcl::make_solver<
    amgcl::amg<
        Backend,
        amgcl::runtime::coarsening::wrapper,
        amgcl::runtime::relaxation::wrapper
        >,
    amgcl::runtime::solver::wrapper<Backend>
    > Solver;

boost::property_tree::ptree prm;

prm.put("solver.type", "bicgstab");
prm.put("solver.tol", 1e-6);
prm.put("precond.coarsening.type", "smoothed_aggregation");
prm.put("precond.relax.type", "spai0");

Solver solve(A, prm);
\end{lstlisting}

\subsection{Extensibility}

Compile-time strategy-based design of the library allows its maintainers and
users to customize the existing components and introduce new functionality by
creating new classes and passing those as template parameters
(\emph{strategies}) to AMGCL templated algorithms. Also, the unified
initialization interface of AMGCL building blocks means it is easy to reuse
existing functionality in order to come up with new preconditioning techniques.
Notable examples included into the library codebase are
CPR~\cite{gries2014preconditioning} and Schur complement pressure
correction~\cite{verfurth1984combined, gmeiner2016quantitative}
preconditioners.  CPR works best in fully implicit black-oil simulations, and
Schur complement pressure correction is well-suited for Navier--Stokes-like
problems. Both of these are field-split, two-step type preconditioners,
allowing to efficiently solve problems arising from discretizations of coupled
systems of PDEs, which are otherwise converge slowly or do not converge at all
if treated as monolithic systems. The pressure subblock of the system matrix is
preconditioned with AMG, and both preconditioners reuse core components of the
library.

\section{Performance and scalability} \label{sec:benchmarks}

In this section the performance and scalability of the shared and distributed
memory versions of the library algorithms are tested using two sample problems
in the three dimensional space. The source code for the benchmarks is available
at https://github.com/ddemidov/amgcl\_benchmarks.

The first example is a classical 3D Poisson problem:
\begin{equation} \label{eq:poisson}
    -\Delta u = 1
\end{equation}
defined in the unit cube $\Omega = \left[0,1\right]^3$ with homogeneous
Dirichlet boundary conditions $\left. u=0 \right|_{\partial\Omega}$. The
problem is discretized using finite difference method on a uniform structured
mesh.

The second example is an incompressible 3D Navier--Stokes problem:
\begin{subequations} \label{eq:ns}
\begin{gather}
    \frac{\partial \mathbf u}{\partial t} + \mathbf u \cdot \nabla \mathbf u  +
    \nabla p = \mathbf b, \\
    \nabla \cdot \mathbf u = 0.
\end{gather}
\end{subequations}
The problem is discretized using an equal-order tetrahedral finite elements,
stabilized employing an ASGS-type (algebraic subgrid-scale)
approach~\cite{Donea2003}. This results in a discretized linear system of
equations with a block structure of the type
\begin{equation} \label{eq:discrete_nstokes}
\begin{bmatrix}
    \mathbf K & \mathbf G \\
    \mathbf D & \mathbf S
\end{bmatrix}
\begin{bmatrix}
    \mathbf u \\ \mathbf p
\end{bmatrix}
=
\begin{bmatrix}
    \mathbf b_u \\ \mathbf b_p
\end{bmatrix},
\end{equation}
where each of the matrix subblocks is a large sparse matrix, and the blocks
$\mathbf G$ and $\mathbf D$ are non-square.  The overall system matrix for the
problem was assembled in the Kratos\footnote{http://www.cimne.com/kratos}
Multi-Physics package~\cite{Dadvand2010,Dadvand2013} developed at CIMNE,
Barcelona.

There are at least two ways to solve the Navier--Stokes problem.  First, one
can treat the system as a monolithic one, and provide some minimal help to the
preconditioner by supplying it with the near null space vectors. Second option
is to employ the knowledge about the problem structure, and to combine separate
preconditioners for individual fields (in this particular case, for pressure
and velocity). In case of AMGCL both options are available and were tested: the
monolithic system was solved with static $4 \times 4$ matrices as value type,
and the field-split approach was implemented using the Schur complement
pressure correction preconditioner, where the pressure part of the system was
preconditioned with smoothed aggregation AMG, and the velocity part was
preconditioned with SPAI-0. Trilinos~ML only provides the first option; PETSC
implements both, but we only show results for the second, superior option of
field-split preconditioner. CUSP library does not provide field-split
preconditioner and nor does it allow to specify near null space vectors, so it
was not tested with the Navier--Stokes problem.

\subsection{Shared memory benchmarks}

The performance of the shared memory (single compute node) solution was tested
on a dual socket system with two Intel Xeon E5-2640 v3 CPUs. The system also
had an NVIDIA Tesla K80 GPU installed, which was used for testing the GPU based
versions. The results were compared to those of the PETSC~\cite{petsc-user-ref}
and the Trilinos~ML~\cite{ml-guide} distributed memory libraries and the CUSP
GPGPU library~\cite{Cusp}.

\begin{figure}
    \begin{center}
        \includegraphics[width=\textwidth]{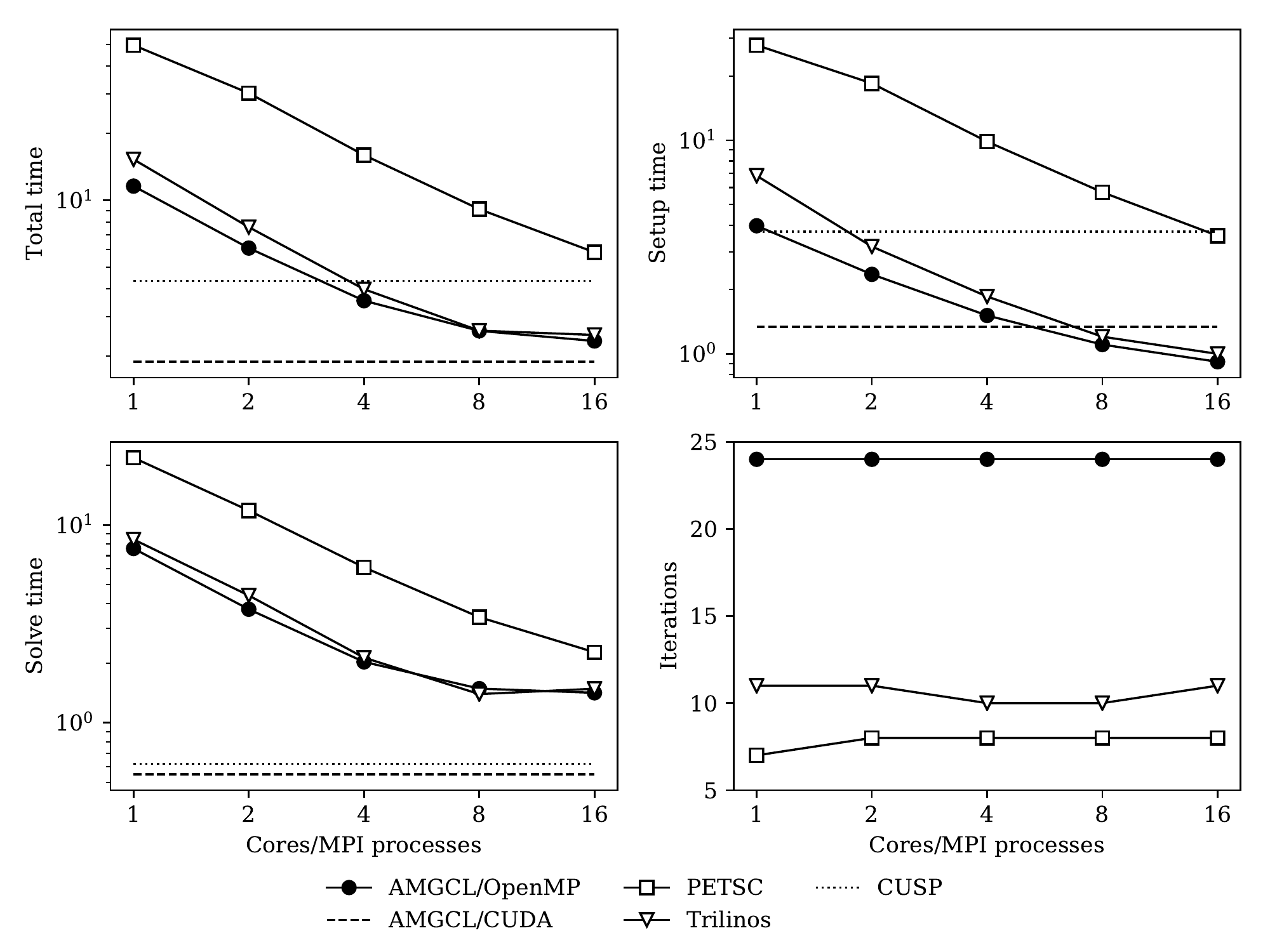}
    \end{center}
    \caption{Shared memory solution of the Poisson problem~\eqref{eq:poisson}.
    The system matrix has $150^3$ unknowns.}
    \label{fig:smem:poisson}
\end{figure}

Figure~\ref{fig:smem:poisson} presents the multicore scalability of the Poisson
problem~\eqref{eq:poisson}. The problem was discretized on a $150^3$ uniform
mesh and the resulting system matrix contained $3\,375\,000$ unknowns and
$23\,490\,000$ nonzero elements. The figure has separate plots for the total
solution time, the setup step time, the solve step time, and the number of
iterations made until convergence. Here PETSC and Trilinos versions use MPI for
parallelisation within the single compute node, AMGCL/OpenMP and AMGCL/CUDA use
OpenMP and CUDA backends correspondingly. Both AMGCL/CUDA and CUSP versions
employ the NVIDIA Tesla K80 GPU. All libraries use the conjugate gradient (CG)
iterative method preconditioned with the smoothed aggregation AMG. Trilinos and
PETSC use their corresponding defaults for smoothers (symmetric Gauss--Seidel
and damped Jacobi accordingly), AMGCL uses SPAI-0, and CUSP uses
Gauss--Seidel.

The CPU-based results show that AMGCL performs on par with Trilinos, and both
outperform PETSC by a large margin. Also, AMGCL is able to setup the solver
about~20--100\% faster than Trilinos, and~4--7 times faster than PETSC. This is
probably due to the fact that both Trilinos and PETSC target distributed memory
machines and thus have an overhead not present in the shared memory version
of AMGCL.

The CUDA backend of AMGCL during solution phase performs slightly better than
the CUSP version, and the setup time for AMGCL is significantly lower than that
of CUSP. The main difference between the libraries is that AMGCL constructs the
AMG hierarchy completely on the CPU, and the setup step in CUSP is done on the
GPU. When taking into consideration just the solution time (without setup),
then both AMGCL/CUDA and CUSP are able to outperform CPU-based versions (with
full utilization of the CPU cores) by a factor of~3--4. The total solution time
for AMGCL/CUDA is only~30\% better than that of AMGCL/OpenMP or Trilinos. This
is explained by the fast convergence rate of the Poisson problem, which means
the solution step of AMGCL/CUDA is not able to amortize the relatively
expensive setup step.

\begin{figure}
    \begin{center}
        \includegraphics[width=\textwidth]{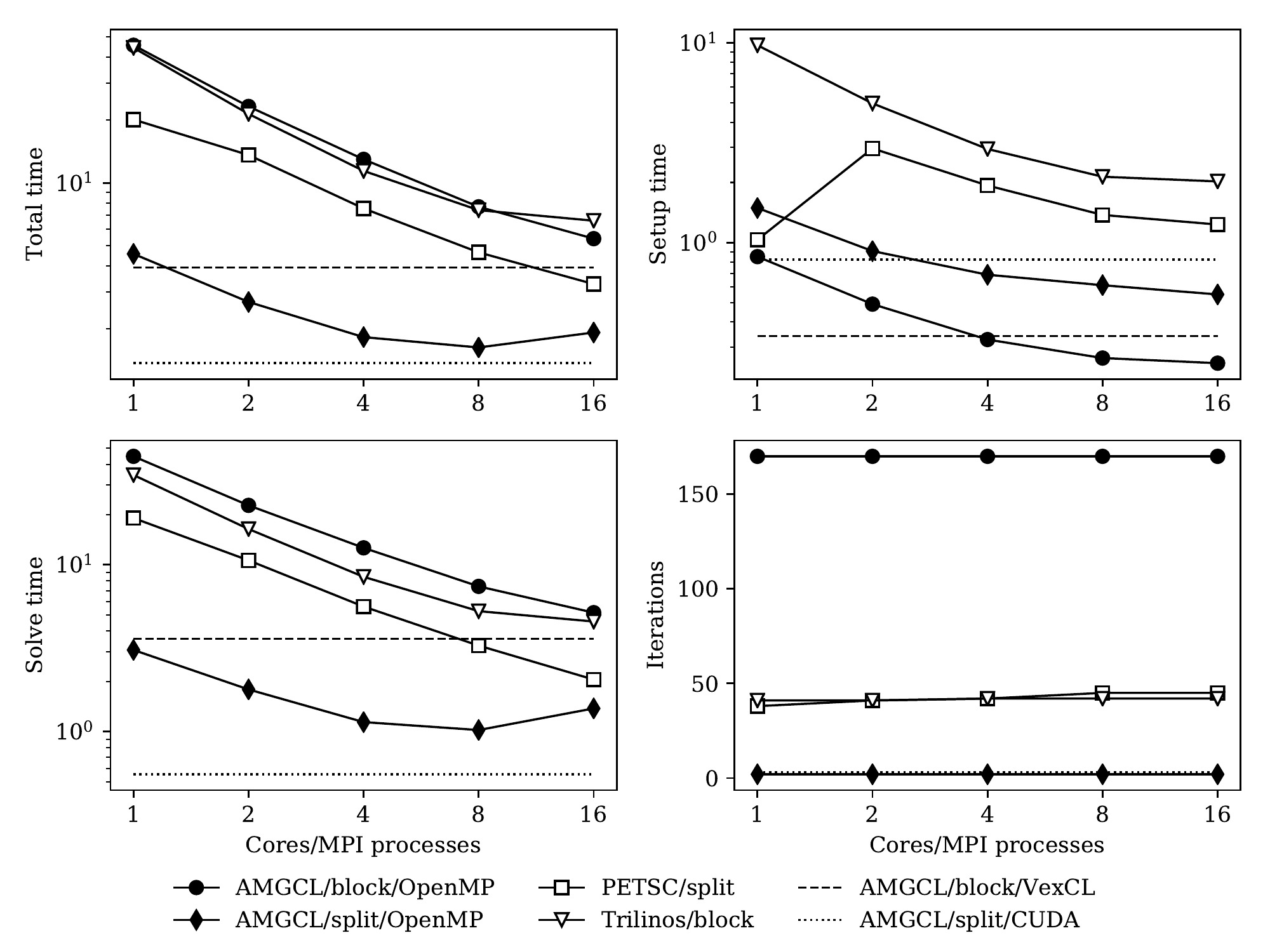}
    \end{center}
    \caption{Shared memory solution of the Navier--Stokes problem~\eqref{eq:ns}.
    The system matrix has $713\,456$ unknowns.}
    \label{fig:smem:ns}
\end{figure}

Figure~\ref{fig:smem:ns} shows the results for the Navier--Stokes
problem~\eqref{eq:ns}. The system matrix has $713\,456$ unknowns and
$41\,277\,920$ nonzeros. The assembled problem files may be downloaded
at~\cite{ns_set_small}. Here the `block' versions correspond to the solution of
the monolithic system with a smoothed aggregation AMG as preconditioner. The
`split' versions correspond to using a field-split preconditioner better suited
for the Navier--Stokes problem. The `split' variants of both AMGCL and PETSC
are predictably able to outperform the `block' versions. AMGCL/block version
with the OpenMP backend performs on par with the Trilinos library, and
AMGCL/split works significantly faster than PETSC. The GPGPU versions of AMGCL
(the VexCL\footnote{https://github.com/ddemidov/vexcl} backend uses OpenCL and
was tested with the `block' preconditioner, and the CUDA version uses `split'
preconditioner), relatively to their CPU-based counterparts, perform similar to
what was observed during solution of the Poisson problem.

\subsection{Distributed memory benchmarks}

The scalability of the distributed memory (MPI based) solutions was tested on a
SuperMUC cluster located at the Leibniz Supercomputing Centre of the Bavarian
Academy of Sciences and Humanities.  Each compute node was equipped with two~14
core Intel Haswell Xeon E5-2697 v3 CPUs, and~64 GB of RAM. The obtained results
were compared to those of the Trilinos~ML package.

\begin{figure}
    \begin{center}
        \subfigure[Weak scaling. Approximately $100^3$ unknowns per MPI process.]{
            \includegraphics[width=0.95\textwidth]{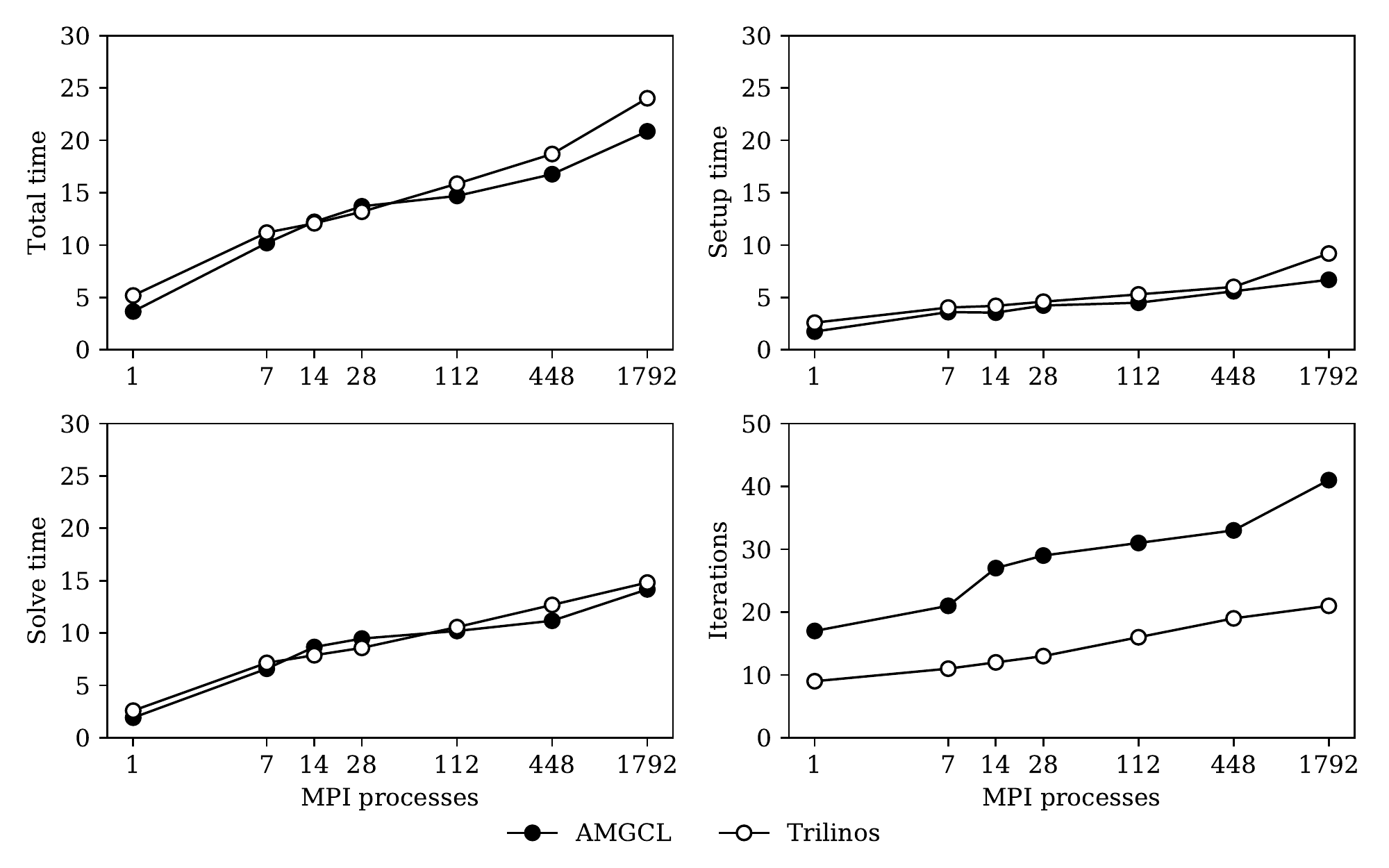}
            \label{fig:dmem:poisson:weak}
        }\\
        \subfigure[Strong scaling. The system matrix has $256^3$ unknowns.]{
            \includegraphics[width=0.95\textwidth]{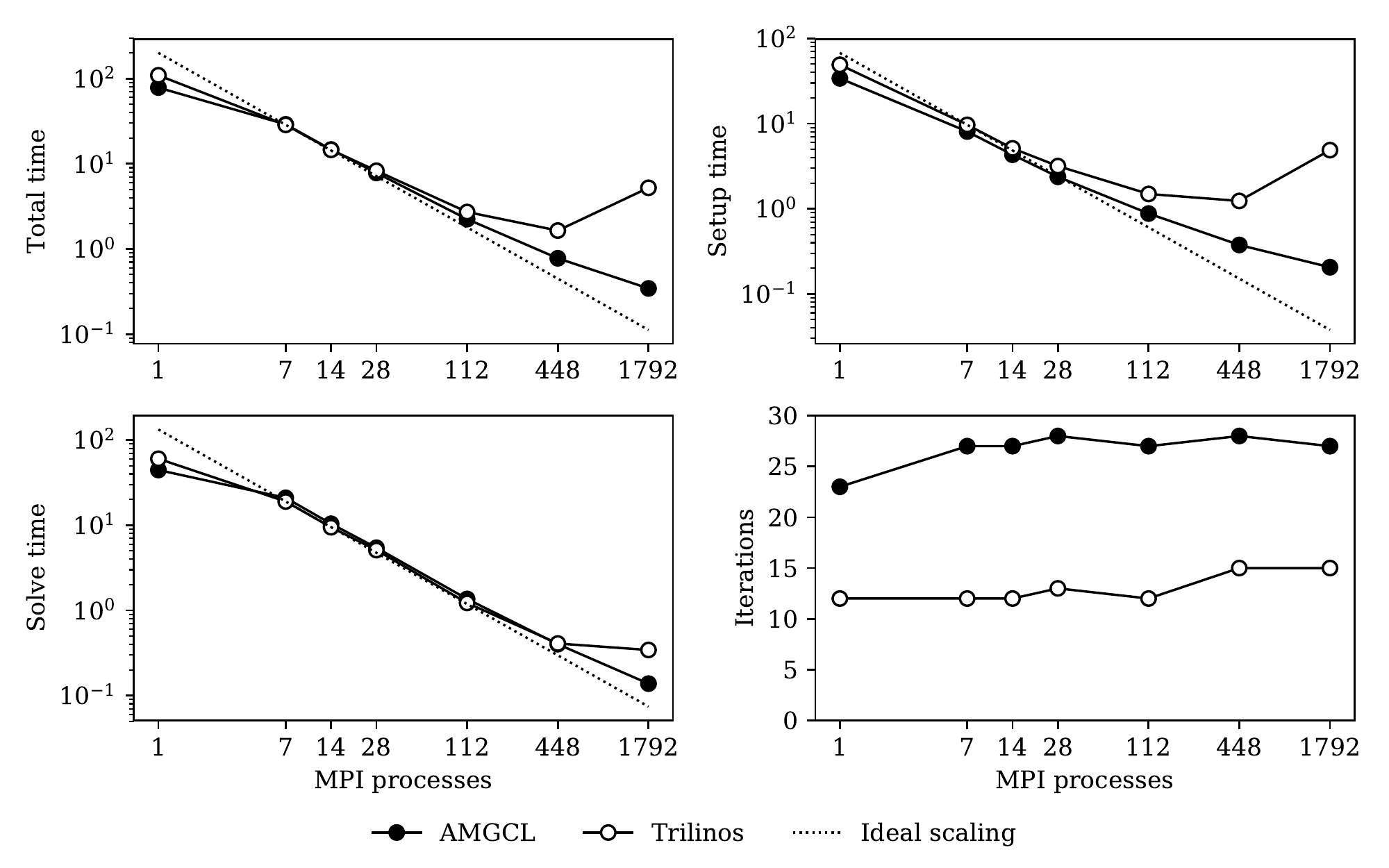}
            \label{fig:dmem:poisson:strong}
        }
    \end{center}
    \caption{Scaling of the Poisson problem~\eqref{eq:poisson} on the SuperMUC
    cluster.}
    \label{fig:dmem:poisson}
\end{figure}

Figure~\ref{fig:dmem:poisson} shows both weak and strong scalability tests of
the Poisson problem solution on the SuperMUC cluster. Weak scaling results are
presented on Figure~\ref{fig:dmem:poisson:weak}.  Here the problem size grows
proportionally to the number of MPI processes so that each process owns
approximately $100^3$ unknowns.  As in the shared memory tests, both AMGCL and
Trilinos versions use conjugate gradient method preconditioned with smoothed
AMG. Trilinos uses symmetric Gauss--Seidel, and AMGCL uses SPAI-0 for
smoothing on each level of the AMG hierarchy. In case of ideal weak scaling the
total solution time should be invariant with respect to the number of MPI
processes. However, both Trilinos and AMGCL demonstrate around 20\% weak
scaling efficiency (defined as $T_1/T_{1792}$) at 1792 MPI processes. Overall,
both libraries show very similar weak scalability.

Figure~\ref{fig:dmem:poisson:strong} depicts strong scalability of the Poisson
problem. Here the problem size is fixed at $256^3$ unknowns. In case of ideal
scaling (presented as dotted line on the figure) the solution time should
decrease with increasing number of MPI processes. Here AMGCL demonstrates
better scalability starting at about 100-400 MPI processes.

\begin{figure}
    \begin{center}
        \includegraphics[width=\textwidth]{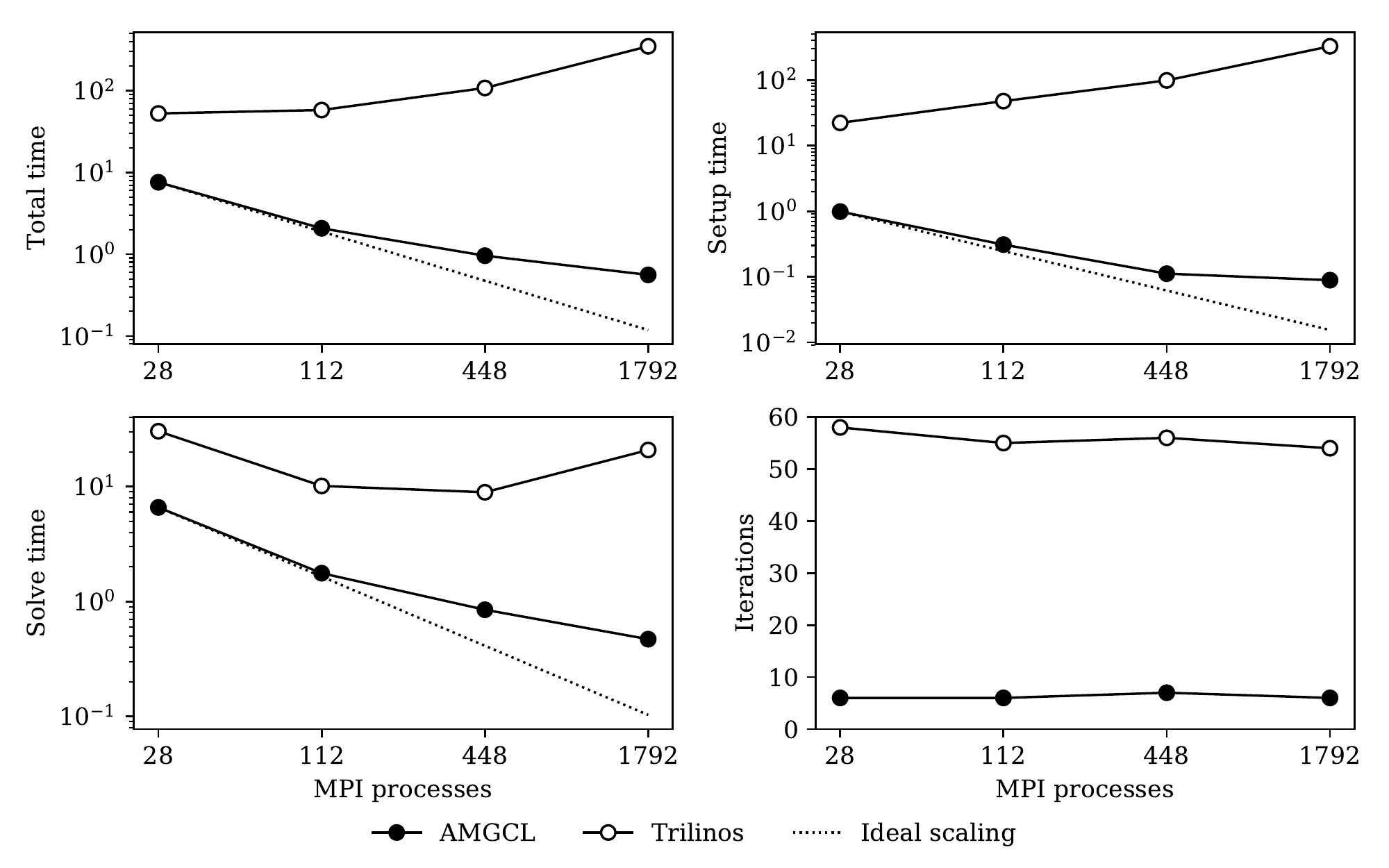}
    \end{center}
    \caption{Scaling of the Navier--Stokes problem~\eqref{eq:ns} on the SuperMUC
    cluster. The system matrix has $4\,773\,588$ unknowns.}
    \label{fig:dmem:ns}
\end{figure}

Scalability of the distributed memory Navier--Stokes solution is presented on
Figure~\ref{fig:dmem:ns}. The system matrix here has $4\,773\,588$ unknowns and
$281\,089\,456$ nonzeros. The assembled problem is available for download
at~\cite{ns_set_large}. The fact that the problem size is fixed means that this
is basically a strong scalability test. AMGCL version uses field split approach
with Schur complement pressure correction preconditioner. Trilinos~ML uses the
nonsymmetric smoothed aggregation variant (NSSA) recommended by the user manual
for this type of problems~\cite{ml-guide}. The tests use default NSSA
parameters. AMGCL scales much better for the Navier--Stokes problem, which is
mostly due to the better choice of preconditioner rather than any omissions in
Trilinos implementation.

\section{Conclusion}

AMGCL is a header only \CXX library that provides an efficient, scalable,  and
customizable implementation of algebraic multigrid method for solution of large
sparse linear systems of equations arising from discretization of partial
differential equations on structured or unstructured grids. The library
supports parallelization with modern hardware using OpenMP, OpenCL, or CUDA
technologies, and may work in shared or distributed memory mode.

The paper provides an overview of design decisions behind the library and
demonstrates the library performance and scalability in comparison with PETSC,
Trilinos~ML, and CUSP packages. Numerical experiments presented in
Section~\ref{sec:benchmarks} of the paper show that AMGCL performs on par or
better than the alternatives on the examples of Poisson and Navier--Stokes
problems.

Among AMGCL advantages are the liberal MIT license it is published under,
possibility to customize, extend, and reuse the library components, and easy
adoption to user data types, which makes integration of AMGCL into existing
projects straightforward. Examples of projects (known to the author) that are
currently using AMGCL are the Kratos Multi-Physics package~\cite{Dadvand2010}
developed at CIMNE, Barcelona, and the MATLAB Reservoir Simulation Toolbox
(MRST)~\cite{lie2014introduction} developed by the Computational Geosciences
group in the  Department of Mathematics and Cybernetics  at SINTEF Digital.
Kratos uses AMGCL as the default linear solver, and MRST provides a MATLAB
interface to linear AMGCL solvers preconditioned with AMG or CPR.

The library source code is available for download at
https://github.com/ddemidov/amgcl.

\section{Acknowledgments}

The development of the AMGCL library was partially funded by the state
assignment to the Joint supercomputer center of the Russian Academy of sciences
for scientific research. Work on field-split type preconditioners was partially
funded by the RFBR grant Nos 18-07-00964, 18-47-160010. Access to the SuperMUC
cluster was provided with support of the PRACE program, project 2010PA4512.

\end{document}